\catcode`\@=11  
\def\answ{p}    
\def\prepri{p }      

\if\answ\prepri
  \documentstyle[11pt,epsf]{article}
\else
   \documentstyle[epsf]{article}
\def\abstract#1{\vskip 7mm
        \begin{center}{\large Abstract}\par \smallskip
                \begin{minipage}[c]{12cm}
                        \small #1
                \end{minipage}
        \end{center}
}
\def\title#1{\begin{center}{\Large\bf #1}\end{center}}
\def\author#1{\vskip 5mm \begin{center}{#1}\end{center}}

\fi

\def\be{\begin{equation}}
\def\en{\end{equation}}
\def\bear{\begin{eqnarray}}
\def\enar{\end{eqnarray}}
\def\beas{\begin{eqnarray*}}
\def\enas{\end{eqnarray*}}
\def\bera{ \setcounter{enumi}{\value{equation}}
           \addtocounter{enumi}{1}
           \setcounter{equation}{0}
           \renewcommand{\theequation}{\theenumi\alph{equation}}
           \begin{eqnarray} }
\def\enra{ \end{eqnarray}
           \setcounter{equation}{\value{enumi}}
           \renewcommand{\theequation}{\arabic{equation}}  }

\def\buildchar#1#2#3{\null \! \mathop {\vphantom {#1}
\smash #1}\limits ^{#2}_{#3}\!\null }
\def\ut#1{\buildchar{#1}{}{^\sim}\/}

\def\been{\begin{enumerate}}
\def\enen{\end{enumerate}}
\def\beit{\begin{itemize}}
\def\enit{\end{itemize}}

\def\bece{\begin{center}}
\def\ence{\end{center}}
\def\bert{\begin{flushright}}
\def\enrt{\end{flushright}}

\def\det{ \mbox{det} \,}

\newtheorem{Def}{Definition}
\begin{document}
\small
\renewcommand{\thefootnote}{\fnsymbol{footnote}}

\if\answ\prepri
\setcounter{footnote}{0}
\else
\setcounter{footnote}{1}
\fi

\begin{center}
~\\~\\
{\Large{\bf
Will hyperbolic formulations help numerical relativity? }}\\
{\large{\bf
--  Experiments using Ashtekar's connection variables --}
\if\answ\prepri
\footnote{
Proceedings of {\it The 10th Workshop on
General Relativity and Gravitation} at Osaka, Japan, September 2000.\\
This report is based on our works \cite{Paper1} and \cite{Paper2}.
}
\fi
}

\vskip .2in

{\large
{\sc  Hisa-aki Shinkai$^{\dagger}$}
and
{\sc  Gen Yoneda$^{\ddagger}$}
}
\\[1em]
{\tt shinkai@gravity.phys.psu.edu} $\quad$
{\tt yoneda@mn.waseda.ac.jp}
\\
$~^{\dagger}$ {\em
Centre for Gravitational Physics and Geometry,
104 Davey Lab., Department of Physics, \\
The Pennsylvania State University,
University Park, Pennsylvania 16802-6300, USA
}\\
$~^{\ddagger}$ {\em Dept. of Mathematical Science, Waseda University,
Shinjuku, Tokyo 169-8555, Japan} 
\\(November 17, 2000 $\quad$ CGPG-00/11-3, gr-qc/0103031)
\end{center}

\abstract{
In order to perform accurate and stable long-term numerical integration
of the Einstein equations, several hyperbolic systems have been proposed.
We here report our numerical comparisons
between weakly hyperbolic, strongly hyperbolic,
and symmetric hyperbolic systems based on Ashtekar's connection variables.
The primary advantage for using this connection formulation 
is that we can keep using the same dynamical variables for all levels of
hyperbolicity.
Our numerical code demonstrates gravitational wave propagation in
plane symmetric spacetimes, and we compare the
accuracy of the simulation by monitoring
the violation of the constraints.
By comparing with results obtained from the weakly hyperbolic system, we
observe the strongly and symmetric hyperbolic system show better
numerical performance (yield less constraint violation),
but not so much difference between the latter two. \\
 We also study asymptotically constrained systems for
numerical integration of the Einstein equations, which are intended to be
robust against  perturbative
errors for the free evolution of the initial data.
First, we examine the previously
proposed ``$\lambda$-system", which introduces artificial flows to
constraint surfaces based on the symmetric hyperbolic formulation.
We show that this system works as expected for the wave propagation
problem in the Maxwell system and
in general relativity using Ashtekar's connection formulation. 
Second, we propose a new mechanism to control the stability, 
which we call 
the ``adjusted system".  This is simply obtained by adding constraint terms
in the dynamical equations and adjusting its multipliers.  We explain
why a particular choice of multiplier reduces the numerical errors
from non-positive or pure-imaginary eigenvalues
of the adjusted constraint propagation equations.
This ``adjusted system" is also tested in the Maxwell system and in the
Ashtekar's system.  This mechanism affects more than the system's
symmetric hyperbolicity.
}


\renewcommand{\thefootnote}{\arabic{footnote})}
\setcounter{footnote}{0}

\section{Introduction} \label{sec:intro}

Numerical relativity -- solving the Einstein equations numerically --
is now an essential field in gravity research.
The current mainstream of numerical relativity is to
demonstrate the final phase of compact
binary objects related to gravitational wave observations \cite{PTPsupple}, 
and these efforts are
now again shedding light on the
mathematical structure of the Einstein equations.


Up to a couple of years ago, the standard Arnowitt-Deser-Misner (ADM) 
decomposition of the Einstein
equations was taken as the standard formulation for numerical relativists.
Difficulties in accurate/stable long-term
evolutions were supposed to be overcome by choosing proper gauge
conditions and boundary conditions.  Recently, however, several
numerical experiments show
that the standard ADM is not the best formulation for numerics, and
finding a better formulation
has become one of the main research topics.

One direction in the community is to apply
conformally decoupled and tracefree re-formulation of ADM
system which were
first used by Nakamura {\it et al.}  \cite{SN}.
Although there is an effort to show why this
re-formulation is better than ADM, 
we do not yet know this method is robust for all situations.

Another alternative approach to ADM is to formulate the 
Einstein equations to reveal 
hyperbolicity \cite{reviewhyp}.
A certain kind of hyperbolicity of the dynamical
equations is essential
to analyze their propagation features mathematically,
and are known to
be useful in numerical approximations.
Several hyperbolic formulations have been proposed to
re-express the Einstein
equations, with different levels:
weakly, strongly and symmetric hyperbolic systems. 
Several numerical tests were also performed in this
direction. 

The following questions, therefore, naturally present themselves
(cf. \cite{Stewart}):
\beit
\item[(1)]
Does hyperbolicity actually contribute to the numerical accuracy/stability?
\item[(2)]
 If so,
which level of hyperbolic formulation is practically useful for numerical
applications? (or does the symmetric hyperbolicity solve all the
difficulties?)
\item[(3)]
 Are there any other approaches to improve the accuracy/stability
  of the system?
\enit


In this report, we try to answer these questions with our
simple numerical experiments.
Such comparisons are appropriate when the fundamental equations
are cast in the same interface, and that is possible 
using Ashtekar's connection variables
\cite{Ashtekar,AshtekarBook}. 
More precisely, the authors' recent studies showed the following:
\beit
\item[(a)]
 the original set of dynamical equations proposed
by Ashtekar already forms 
a weakly hyperbolic system \cite{YS-IJMPD},
\item[(b)]
 by requiring additional gauge conditions {\it or} adding constraints to
the dynamical equations, we can obtain a
strongly hyperbolic system \cite{YS-IJMPD},
\item[(c)]
by requiring additional gauge conditions {\it and} adding constraints to
the dynamical equations,  we can obtain a symmetric hyperbolic system
\cite{YS-IJMPD,YShypPRL}, and finally
\item[(d)]
 based on the above symmetric hyperbolic system,
we can construct a set of
dynamical systems which is robust against perturbative
errors for constraints and
reality conditions \cite{SY-asympAsh}
({\it aka.} $\lambda$-system \cite{BFHR}).
\enit

Based on the above results (a)-(c),
we developed a numerical code which handles
gravitational wave propagation in the
plane symmetric spacetime.
We performed the time evolutions using
  the above three levels of Ashtekar's dynamical equations
  together with the standard ADM equation.
We compare these for accuracy and stability
by monitoring the violation of the constraints.
We also show the demonstrations of our $\lambda$-system (above (d)), 
together with new proposal for controlling the stability.

It is worth remarking that this study is the first one which shows
full numerical simulations of Lorentzian spacetime using
Ashtekar's connection variables.
This research direction was suggested
\cite{AshtekarRomano89} soon after Ashtekar completed
his formulation, but has not yet been completed.
Historically, an application to numerical relativity of the connection
formulation was also suggested \cite{AshtekarBook,Salisbury}
using Capovilla-Dell-Jacobson's version of the connection variables
\cite{CDJ}, which produce an direct relation to Newman-Penrose's $\Psi$s.
However here we apply Ashtekar's original formulation, because we know
how to treat its reality conditions in detail, 
and how they form hyperbolicities. 
We will also describe the basic numerical procedures in this paper.

Due to the limited space, we have to omit details in some part in this 
report. More complete discussion can be found in our recent
articles \cite{Paper1,Paper2}. 

\section{Field equations to be compared}
\subsection{Hyperbolic formulations}
We begin by providing our definitions of
hyperbolic systems.
\def\cha{J}
\def\coe{K}
\begin{Def}
We assume
a certain set of
(complex) variables $u_\alpha$ $(\alpha=1,\cdots, n)$
forms a first-order (quasi-linear)
partial differential equation system,
\begin{equation}
\partial_t u_\alpha
= \cha ^{l}{}^{\beta}{}_\alpha (u) \, \partial_l u_\beta
+\coe_\alpha(u),
\label{def}
\end{equation}
where $\cha$ (the characteristic matrix) and
$\coe$ are functions of $u$
but do not include any derivatives of $u$.
We say that the system (\ref{def}) is:
\been
\def\theenumi{(\Roman{enumi})}
\item  \label{weakhyp}
{\bf weakly hyperbolic},
if all the eigenvalues of the
characteristic matrix are real. 
\item \label{diaghyp}
{\bf strongly (diagonalizable) hyperbolic},
if the characteristic matrix is
diagonalizable and has all real eigenvalues.
\item \label{symhyp}
{\bf symmetric hyperbolic},
if the characteristic matrix is a
Hermitian matrix. 
\enen
\end{Def}
We treat $\cha^{l\beta}{}_\alpha$ as a $n \times n$ matrix
when  the $l$-index is fixed.
We say $\lambda^l$ is an
eigenvalue of $\cha^{l\beta}{}_\alpha$
when the characteristic equation,
$\det (\cha^{l\beta} {}_\alpha
-\lambda^l \delta^\beta {}_\alpha)=0$,
is satisfied.
The eigenvectors, $p^\alpha$, are given by solving
$\cha^{l}{}^{\alpha} {}_\beta \, p^{l\beta}=\lambda^l  p^{l\alpha}$.

These three classes have the relation
(III)  $\in$ (II) $\in$ (I).
The symmetric hyperbolic system gives us the energy
integral inequalities,
which are the primary tools for studying
well-posedness of the system.

For more concrete descriptions for each systems, please
refer our paper \cite{YS-IJMPD}.
It might be worth remarking that
the standard ADM formulation cannot be a first-order
hyperbolic form, since there are curvature terms in the r.h.s. of
dynamical equations.

\subsection{Ashtekar's dynamical equations: three levels of hyperbolicity}
The key feature of  Ashtekar's formulation of general relativity
\cite{Ashtekar} is the introduction of a self-dual
connection as one of the basic dynamical variables.
The new basic variables are
the densitized inverse triad, $\tilde{E}^i_a$, and the 
SO(3,C) self-dual connection, ${\cal A}^a_i$, where the indices
$i,j,\cdots$ indicate the 3-spacetime, and
$a,b,\cdots$ are for SO(3) space.
The total four-dimensional spacetime is described together with the
gauge variables
$\null \! \mathop {\vphantom {N}\smash N}\limits ^{}_{^\sim}\!\null
, N^i, {\cal A}^a_0$, which we call the densitized lapse
function, shift vector and the triad lapse function.
The system has three constraint equations,
\begin{eqnarray}
{\cal C}^{\rm ASH}_{H} &:=&
  (i/2)\epsilon^{ab}{}_c \,
\tilde{E}^i_{a} \tilde{E}^j_{b} F_{ij}^{c}
    \approx 0, \label{const-ham} \\
{\cal C}^{\rm ASH}_{M i} &:=&
   -F^a_{ij} \tilde{E}^j_{a} \approx 0, \label{const-mom}\\
{\cal C}^{\rm ASH}_{Ga} &:=&  {\cal D}_i \tilde{E}^i_{a}
  \approx 0,  \label{const-g}
\end{eqnarray}
which are called the Hamiltonian, momentum, and Gauss constraints equation,
respectively.
The dynamical equations for a set of
$(\tilde{E}^i_a, {\cal A}^a_i)$ are
\begin{eqnarray}
\partial_t {\tilde{E}^i_a}
&=&-i{\cal D}_j( \epsilon^{cb}{}_a  \, \null \!
\mathop {\vphantom {N}\smash N}\limits ^{}_{^\sim}\!\null
\tilde{E}^j_{c}
\tilde{E}^i_{b})
+2{\cal D}_j(N^{[j}\tilde{E}^{i]}_{a})
+i{\cal A}^b_{0} \epsilon_{ab}{}^c  \, \tilde{E}^i_c,  \label{eq-E}
\\
\partial_t {\cal A}^a_{i} &=&
-i \epsilon^{ab}{}_c  \,
\null \! \mathop {\vphantom {N}\smash N}\limits ^{}_{^\sim}\!\null
\tilde{E}^j_{b} F_{ij}^{c}
+N^j F^a_{ji} +{\cal D}_i{\cal A}^a_{0},
\label{eq-A}
\end{eqnarray}
where
${F}^a_{ij}
:=
2 \partial_{[i} {\cal A}^a_{j]}
  - i \epsilon^{a}{}_{bc} \, {\cal A}^b_i{\cal A}^c_j
$
is the curvature 2-form.

We have to consider the reality conditions when we use this
formalism to describe the classical Lorentzian spacetime.
Fortunately, 
the metric will remain on
its real-valued constraint surface during time evolution
  automatically if we prepare initial data which satisfies the
reality condition. 
More practically, we further require that triad is
real-valued.  But again this reality condition appears as a gauge
restriction on ${\cal A}^a_0$\cite{ys-con}, 
which can be imposed at every time step. In our actual simulation, we
prepare our initial data using the standard ADM approach, so that
we have no difficulties in maintaining  these reality conditions.

The set of
dynamical equations (\ref{eq-E}) and (\ref{eq-A})
 [the {\it original} equations] 
does have
a weakly hyperbolic form \cite{YS-IJMPD}, so that we regard  
the mathematical structure of the original equations as one step
advanced from the standard ADM.  
Further, we can construct higher levels of
hyperbolic systems by restricting the gauge condition and/or
by adding constraint terms,
${\cal C}^{\rm ASH}_{H}, {\cal C}^{\rm ASH}_{Mi}$ and
${\cal C}^{\rm ASH}_{Ga}$, to the original equations
\cite{YS-IJMPD}.  We extract only the final expressions here.

In order to obtain a
symmetric hyperbolic system
\footnote{
Iriondo et al \cite{Iriondo} presented 
a symmetric hyperbolic expression 
in a different form.
The differences between ours and theirs are discussed in
\cite{YS-IJMPD,YShypPRL}.}, we add constraint terms to
the right-hand-side of
(\ref{eq-E}) and (\ref{eq-A}).  The adjusted dynamical equations, 
\begin{eqnarray}
\partial_t {\tilde{E}^i_a}
&=&-i{\cal D}_j( \epsilon^{cb}{}_a  \, \null \!
\mathop {\vphantom {N}\smash N}\limits ^{}_{^\sim}\!\null
\tilde{E}^j_{c}
\tilde{E}^i_{b})
+2{\cal D}_j(N^{[j}\tilde{E}^{i]}_{a})
+i{\cal A}^b_{0} \epsilon_{ab}{}^c  \, \tilde{E}^i_c
+\kappa_1 P^i{}_{ab}  \, {\cal C}^{\rm ASH}_G{}^b,  \label{eqE2} \\
&~& {\rm where} \qquad P^i{}_{ab} \equiv
N^i \delta_{ab}+i\null \! \mathop {\vphantom {N}\smash N}
\limits ^{}_{^\sim}\!\null  \epsilon_{ab}{}^{c}\tilde{E}^i_c,
\nonumber
\\
\partial_t {\cal A}^a_{i} &=&
-i \epsilon^{ab}{}_c  \,
\null \! \mathop {\vphantom {N}\smash N}\limits ^{}_{^\sim}\!\null
\tilde{E}^j_{b} F_{ij}^{c}
+N^j F^a_{ji} +{\cal D}_i{\cal A}^a_{0}
+\kappa_2Q^a_i {\cal C}^{\rm ASH}_H
+\kappa_3R_i{}^{ja}  \, {\cal C}^{\rm ASH}_{Mj}, \label{eqA2}
\\
&~& {\rm where} \qquad Q^a_{i} \equiv
e^{-2}
\null \! \mathop {\vphantom {N}\smash N}\limits ^{}_{^\sim}\!\null
\tilde{E}^a_i, \qquad
R_i{}^{ja} \equiv
ie^{-2}
\null \! \mathop {\vphantom {N}\smash N}\limits ^{}_{^\sim}\!\null
\epsilon^{ac}{}_b \tilde{E}^b_i \tilde{E}^j_c
\nonumber
\end{eqnarray}
form a symmetric hyperbolicity if we further 
require $\kappa_1=\kappa_2=\kappa_3=1$ and the gauge conditions,
\begin{equation}
{\cal A}^a_0={\cal A}^a_i N^i, \qquad \partial_i N =0.
\label{symhypgauge}
\end{equation}
We remark that the adjusted coefficients,
$P^i{}_{ab}, Q^a_i, R_i{}^{ja}$, for
constructing the symmetric  hyperbolic system are uniquely determined,
and
there are no other additional terms (say, no ${\cal C}^{\rm ASH}_H,
{\cal C}^{\rm ASH}_M$ for $\partial_t \tilde{E}^i_a$,
no ${\cal C}^{\rm ASH}_G$ for $\partial_t {\cal A}^a_i$)
\cite{YS-IJMPD}.
The gauge conditions, (\ref{symhypgauge}), are
consequences of the consistency with (triad) reality conditions.

We can also construct a strongly hyperbolic system
by restricting to a gauge
$N^l \neq 0, \pm N \sqrt{\gamma^{ll}}$
(where $\gamma^{ll}$ is the three-metric and we do not sum indices here)
for the original equations (\ref{eq-E}), (\ref{eq-A}).
Or we can also construct from the adjusted equations,
(\ref{eqE2}) and (\ref{eqA2}),
  together with the gauge condition
\begin{equation}
{\cal A}^a_0={\cal A}^a_i N^i.  \label{diagohypgauge}
\end{equation}
As for the strongly hyperbolic system, we hereafter take the latter
expression.  


\begin{table}[h]
\begin{center}
\begin{tabular}{cl||l|l|c}
\hline
& system & variables & Eqs of motion & remark
\\
\hline \hline
I & Ashtekar (weakly hyp.) & ($\tilde{E}^i_a, {\cal A}^a_i$) &
(\ref{eq-E}), (\ref{eq-A}) (original) & ``original Ashtekar"
\\ \hline
II & Ashtekar (strongly hyp.) & ($\tilde{E}^i_a, {\cal A}^a_i$) &
(\ref{eqE2}), (\ref{eqA2}) (adjusted with $\kappa=1$)&
(\ref{diagohypgauge}) required
\\ \hline
III & Ashtekar (symmetric hyp.) & ($\tilde{E}^i_a, {\cal A}^a_i$) &
(\ref{eqE2}), (\ref{eqA2}) (adjusted with $\kappa=1$)&
(\ref{symhypgauge}) required
\\ \hline
& Ashtekar (lambda) &
($\tilde{E}^i_a, {\cal A}^a_i, \lambda$s) &
(\ref{DClambda-system}), (\ref{lambdaC1}) - (\ref{lambdaC3})&
\\ \hline
& Ashtekar (adjusted) &
($\tilde{E}^i_a, {\cal A}^a_i$) &
(\ref{eqE2}), (\ref{eqA2}) (adjusted with different $\kappa$)&
\\
\hline
\end{tabular}

\caption{List of systems that we compare in this article.  }
\label{eqmtable}
\end{center}
\end{table}

\subsection{Alternative approaches to obtain robust evolution 1:
``$\lambda$-system"}
Based on the symmetric hyperbolic feature of the system, 
Brodbeck, Frittelli, H\"ubner and Reula (BFHR) proposed an  
alternative dynamical system to obtain stable evolutions, 
which they named 
``$\lambda$-system" \cite{BFHR}.
The idea of this approach is to introduce additional variables, $\lambda$,
which indicates the violation of the constraints, and to construct a symmetric
hyperbolic system for both the original variables and $\lambda$s together with
imposing dissipative dynamical equations for $\lambda$s.
BFHR constructed their $\lambda$-system based on Frittelli-Reula's
symmetric hyperbolic formulation of the Einstein equations \cite{FR96}, and we
\cite{SY-asympAsh} have
also presented the similar system for the Ashtekar's connection
formulation 
based on the above symmetric hyperbolic expression.  Here we present 
our system which evolves the
spacetime to the constraint surface, ${\cal C}_H \approx {\cal C}_{Mi}
\approx {\cal C}_{Ga} \approx 0$ as the attractor.
In \cite{SY-asympAsh}, we also present a system which controls the 
perturvative violation of the reality condition. 

We introduce new variables ($\lambda, \lambda_i, \lambda_a$),
as they obey the dissipative evolution equations
\begin{eqnarray}
\partial_t\lambda &=&
\alpha_1 \,{\cal C}_H
-\beta_1 \, \lambda, \label{lambdaC1}
\\
\partial_t\lambda_i &=&
 \alpha_2 \,\tilde{{\cal C}}_{Mi}
 -\beta_2 \,\lambda_i, \label{lambdaC2}
\\
\partial_t\lambda_a &=&
\alpha_3 \, {\cal C}_{Ga}
-\beta_3 \, \lambda_a, \label{lambdaC3}
\end{eqnarray}
where $\alpha_i \neq 0$ (allowed to be complex numbers) and $\beta_i > 0$
(real numbers) are constants.

If we take
${u}^{(DL)}_\alpha=(\tilde{E}^i_a, {\cal A}^a_i, \lambda, \lambda_i, \lambda_a)$
as a set of dynamical variables, then the
principal part of (\ref{lambdaC1})-(\ref{lambdaC3})
can be written as
\begin{eqnarray}
\partial_t\lambda &\cong&
 -i\alpha_1\epsilon^{bcd} \tilde{E}^j_c \tilde{E}^l_d  (\partial_l{\cal A}^b_j),
\\
\partial_t\lambda_i&\cong&
\alpha_2
[-e \delta^l_i \tilde{E}^j_b
+e \delta^j_i \tilde{E}^l_b
](\partial_l{\cal A}^b_j),
\\
\partial_t\lambda_a&\cong&\alpha_3
\partial_l\tilde{E}^l_a.
\end{eqnarray}

The characteristic matrix of the system ${u}^{(DL)}_\alpha$ does not
form a Hermitian matrix.  However,
if we modify the right-hand-side of
the evolution equation of ($\tilde{E}^i_a, {\cal A}^a_i$), 
then the set becomes
a symmetric hyperbolic system.
This is done by adding
$\bar{\alpha}{}_3 \gamma^{il}(\partial_l \lambda_a)$
to the equation of $\partial_t \tilde{E}^i_a$,
and by adding
$i\bar{\alpha}{}_1\epsilon^a{}_c{}^d \tilde{E}^c_i \tilde{E}^l_d
(\partial_l \lambda)
+
\bar{\alpha}{}_2
(-e \gamma^{lm} \tilde{E}^a_i
+e \delta^m_i \tilde{E}^{la}  )
(\partial_l \lambda_m)
$ to the equation of $\partial_t{\cal A}^a_i$.
The final principal part, then, is written as
\begin{equation}
\partial_t \left(
\matrix{\tilde{E}^i_a \cr {\cal A}^a_i \cr \lambda
 \cr \lambda_i \cr \lambda_a}
\right)
\cong
\left(
\matrix{
A^l {}_a {}^{bi}{}_m & 0 & 0 & 0&
 \bar{\alpha}{}_3 \gamma^{il}\delta_a{}^b  
\cr 
0&  D^l{}^a{}_i{}_b{}^m&
i\bar{\alpha}{}_1\epsilon^a{}_c{}^d \tilde{E}^c_i \tilde{E}^l_d
&
\bar{\alpha}{}_2 e 
(
 \delta^m_i \tilde{E}^{la} - \gamma^{lm} \tilde{E}^a_i )
 & 0 
\cr 
0 & 
-i\alpha_1\epsilon_b{}^{cd} \tilde{E}^m_c \tilde{E}^l_d
& 0 & 0 & 0 
\cr 
0 &
\alpha_2 e 
(\delta^m_i \tilde{E}^l_b -\delta^l_i \tilde{E}^m_b)
& 0 & 0 & 0  
\cr 
\alpha_3\delta_a{}^b \delta^l{}_m& 0 & 0 & 0& 0
}
\right)
\partial_l \left(
\matrix{\tilde{E}^m_b \cr {\cal A}^b_m \cr \lambda
 \cr \lambda_m \cr \lambda_b}
\right).  
\label{DClambda-system}
\end{equation}
Clearly, the solution
$(\tilde{E}^i_a, {\cal A}^a_i, \lambda, \lambda_i, \lambda_a)
=(\tilde{E}^i_a, {\cal A}^a_i, 0, 0, 0)$ represents the original solution
of the Ashtekar system.  If the $\lambda$s decay to zero
after the evolution, then the solution also describes the original
solution of the Ashtekar system in that stage.
Since the dynamical system of ${u}^{(DL)}_\alpha$, 
(\ref{DClambda-system}),
 constitutes a symmetric
hyperbolic form, the solutions to the $\lambda$-system are unique.
BFHR showed analytically that such a decay of $\lambda$s can be seen
for $\lambda$s sufficiently close to zero
with a choice of appropriate combination of $\alpha$s and $\beta$s, 
and that statement can be
also applied to our system.
Therefore, the dynamical system, (\ref{DClambda-system}), is useful
for stabilizing numerical simulations from the point that it recovers
the constraint surface automatically.

\subsection{Alternative approaches to obtain robust evolution 2: 
``adjusted system"}
We also try to compare a set of evolution system, which we propose as
``adjusted-system".
The essential procedures are 
to add constraint terms in the right-hand-side of the dynamical
equations with multipliers, and to choose the multipliers so as to 
decrease the violation of constraint equation.
This second step will be explained by obtaining
non-positive (or non-zero) eigenvalues
of the adjusted constraint propagation equations.
We remark that this eigenvalue criterion is also the core part of 
the theoretical support of the above $\lambda$-system.

The fundamental equations that we will demonstrate in this report,
are the same with (\ref{eqE2}) and (\ref{eqA2}), but here the
real-valued constant multipliers $\kappa$s are not necessary equals to
unity.  We set $\kappa\equiv\kappa_1=\kappa_2=\kappa_3$ for simplicity. 
Apparently 
the set of (\ref{eqE2}) and (\ref{eqA2})
becomes the original weakly hyperbolic system if $\kappa=0$,
becomes the symmetric hyperbolic system if $\kappa=1$ and
$N=const.$. The set remains strongly hyperbolic systems
for other choices of $\kappa$ except $\kappa=1/2$ which only forms
 weakly hyperbolic system.
\section{Comparing numerical performance}
\subsection{Model and Numerical method}
The model we present here is gravitational wave propagation in a
planar spacetime under periodic boundary condition. 
We perform a full numerical simulation using
Ashtekar's variables.  
We prepare two $+$-mode strong pulse waves initially
by solving the ADM Hamiltonian constraint equation, 
using York-O'Murchadha's conformal approach. Then we
transform the initial Cauchy data (3-metric and extrinsic curvature)
into the connection variables, $(\tilde{E}^i_a, {\cal A}^a_i)$,
and evolve them using the dynamical equations. 
For the presentation in this article, we apply the geodesic slicing condition
(ADM lapse $N=1$ or densitized lapse $\ut N=1$, 
with zero shift and zero triad
lapse).  We have used both the Brailovskaya integration scheme, which is
a second order predictor-corrector method, and the so-called iterative 
Crank-Nicholson integration scheme for numerical time evolutions.    
The details of the numerical method are
described in \cite{Paper1}, where we also described how our
code shows second order convergence behaviour. 

 More specifically, we set our initial
guess 3-metric as
\begin{equation}
\hat{\gamma}_{ij}=
\left(\matrix{
1&0&0 \cr
sym.& 1+ K (e^{ -  (x-L)^2}+  e^{ -  (x+L)^2})&0 \cr
sym.& sym.& 1- K (e^{ - (x-L)^2} + e^{ - (x+L)^2})
}\right), 
\label{plusmetric}
\end{equation}
in the periodically bounded region $x=[-5, +5]$.
 Here $K$ and $L$ are constants
and we set $K=0.3$ and $L=2.5$ for the plots. 

In order to show the expected ``stabilization behaviour" clearly,
we artificially add an error in the
middle of the time evolution. We added an artificial
inconsistent rescaling once at time $t=6$ for the ${\cal A}^2_y$ component as
${\cal A}^2_y \rightarrow {\cal A}^2_y (1+ {\rm error})$. 

\subsection{Differences between three levels of hyperbolicity}
We have performed comparisons of stability and/or accuracy between weakly and
strongly hyperbolic systems, and between weakly and symmetric hyperbolic
systems\cite{Paper1}. (We can not compare strongly and symmetric hyperbolic systems
directly, because these two requires different gauge conditions.)

We omit figures in this report, but one can see a part of results in 
Fig.\ref{errwave} and Fig.\ref{errwave2}. 
We may conclude that 
higher level hyperbolic system
gives us slightly accurate evolutions. 
However, if we evaluate the magnitude of L2 norms, then 
we also conclude that there is no measurable differences between 
strongly and symmetric hyperbolicities.  
This last fact will be supported more affirmatively in the next
experiments.

\subsection{Demonstrating ``$\lambda$-system"}
\begin{figure}[h]
\setlength{\unitlength}{1in}
\begin{picture}(6.0,3.0)
\put(0.0,0.25){\epsfxsize=2.8in \epsfysize=1.8in \epsffile{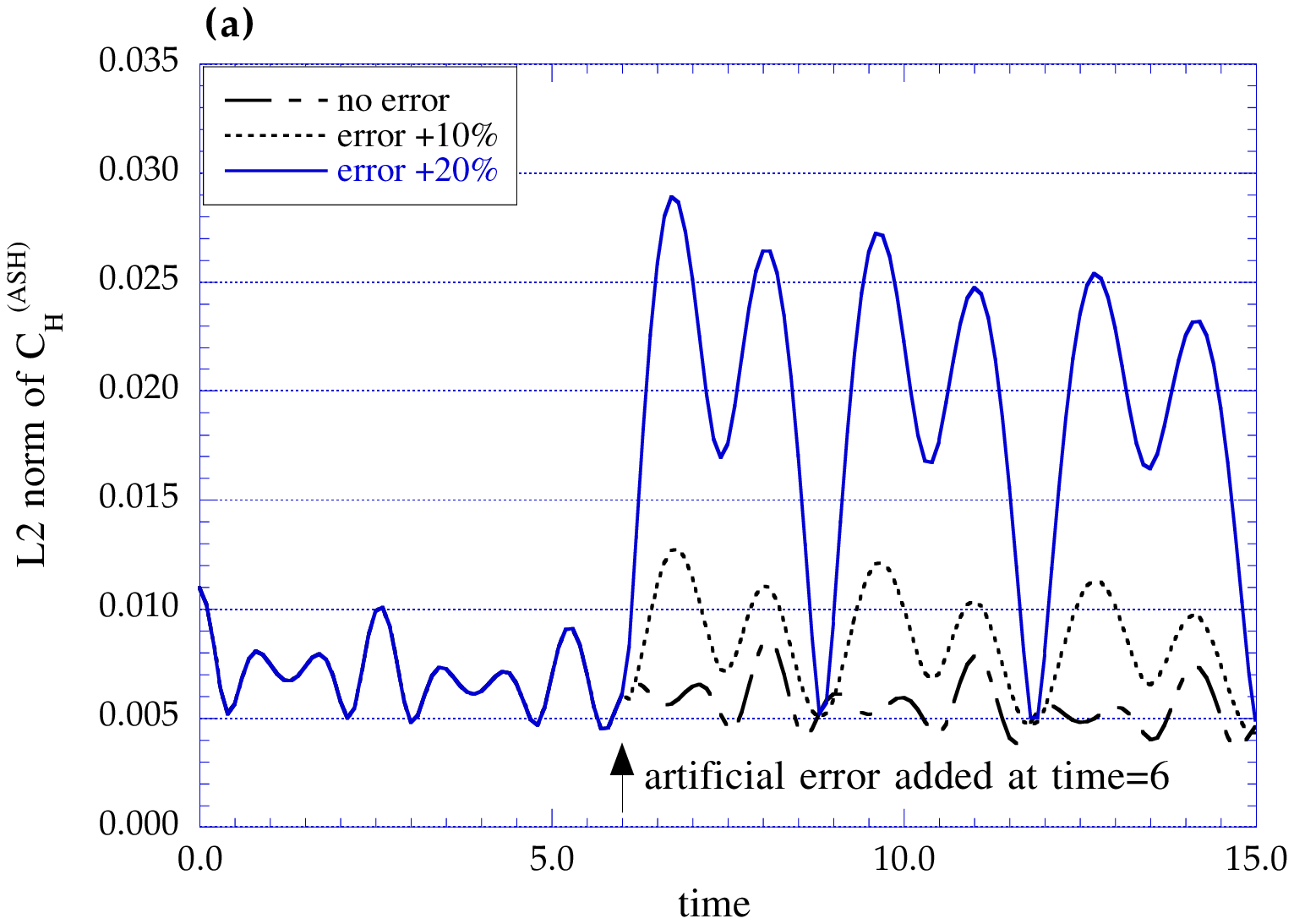} }
\put(3.2,0.25){\epsfxsize=2.8in \epsfysize=1.8in \epsffile{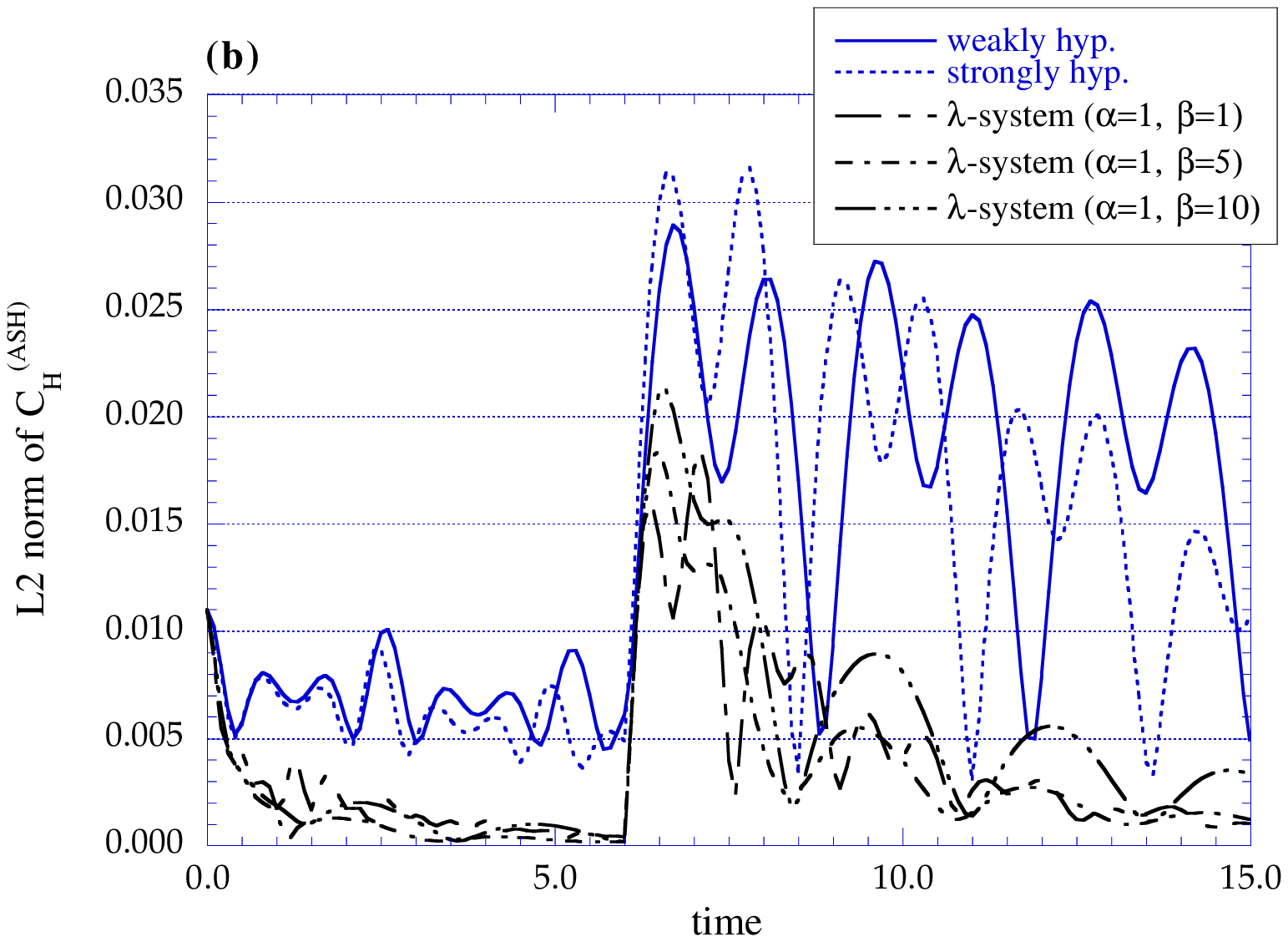} }
\end{picture}

\caption[fig-acc]{\small
Demonstration of the $\lambda$-system in the Ashtekar equation.
We plot the violation of the constraint (L2 norm of the Hamiltonian
constraint equation, ${\cal C}_H$)
for the cases of plane wave propagation under the
periodic boundary.  To see the effect more clearly, we added artificial
error at $t=6$.
Fig. (a) shows how the system goes bad depending on the amplitude of
artificial error.  The error was of the form 
${\cal A}^2_y \rightarrow {\cal A}^2_y (1+ \mbox{ error})$.  
All the lines are of the
evolution of Ashtekar's original equation (no $\lambda$-system).
Fig. (b) shows the effect of $\lambda$-system.
All the lines are 20\% error amplitude, but shows the difference of
evolution equations. The solid line is for Ashtekar's original equation
(the same as in Fig.(a)), the dotted line is for the strongly hyperbolic
Ashtekar's equation.  Other lines are of $\lambda$-systems, which produces
better performance than that of the strongly hyperbolic system.
}
\label{errwave}
\end{figure}

Next, we show a result of the ``$\lambda$-system" \cite{Paper2}.
Fig.\ref{errwave} (a) shows how the violation of the Hamiltonian
constraint equation, ${\cal C}_H$, become  worse depending on the
term ${\rm error}$.
The oscillation of the L2 norm ${\cal C}_H$ in the figure due to the
pulse waves collide periodically in the numerical region.
We, then, fix the error term as a 20\% spike, and try to evolve the
same data in different equations of motion, i.e., the original Ashtekar's
equation [solid line in Fig.\ref{errwave} (b)], strongly hyperbolic
version of Ashtekar's equation (dotted line) and the above $\lambda$-system
equation (other lines) with different $\beta$s but the same $\alpha$.
As we expected, all the $\lambda$-system cases result in reducing the 
Hamiltonian constraint errors.


\subsection{Demonstrating ``adjusted system"}

\begin{figure}[h]
\setlength{\unitlength}{1in}
\begin{picture}(6.0,3.0)
\put(0.0,0.25){\epsfxsize=2.8in \epsfysize=1.8in \epsffile{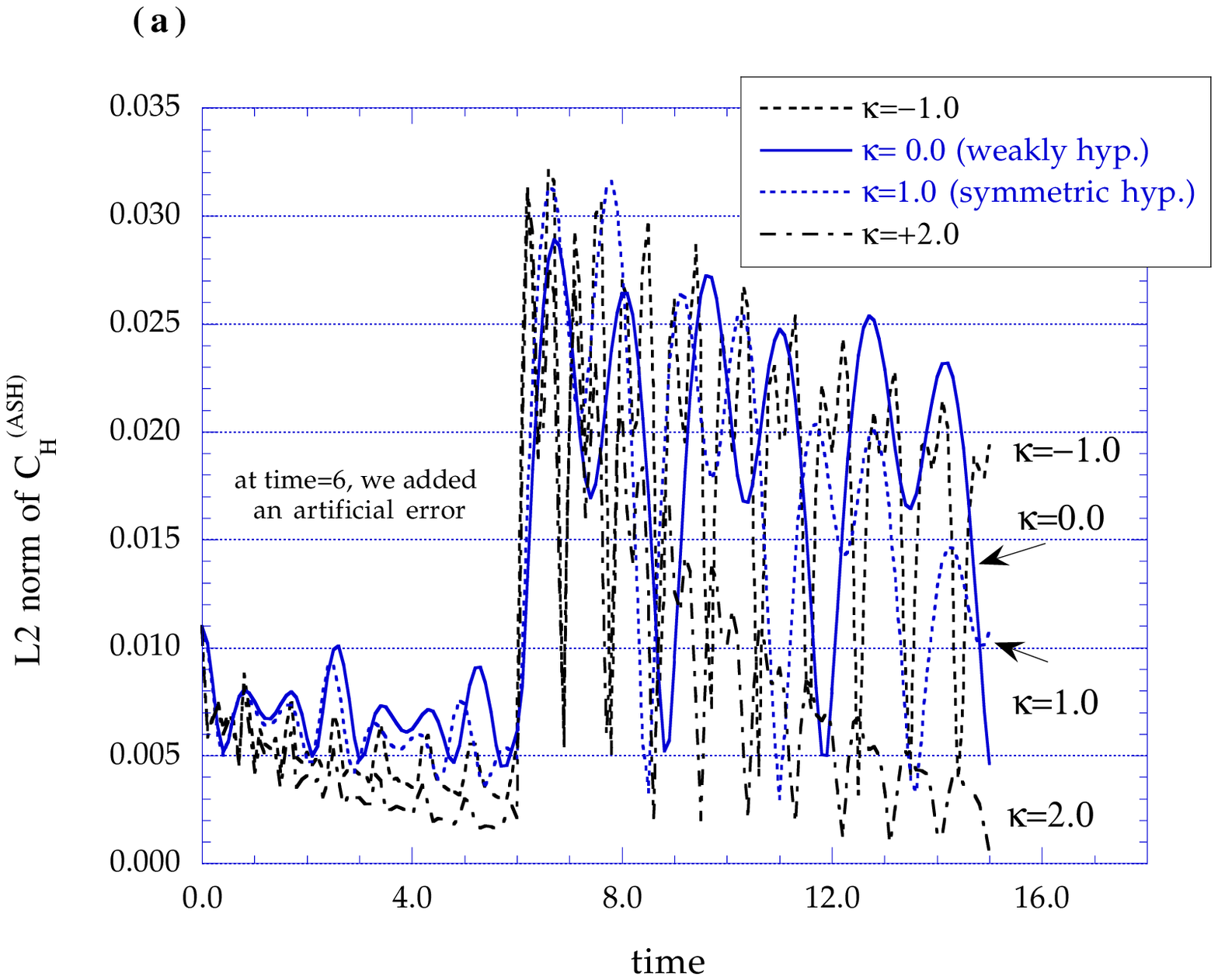} }
\put(3.2,0.25){\epsfxsize=2.8in \epsfysize=1.8in \epsffile{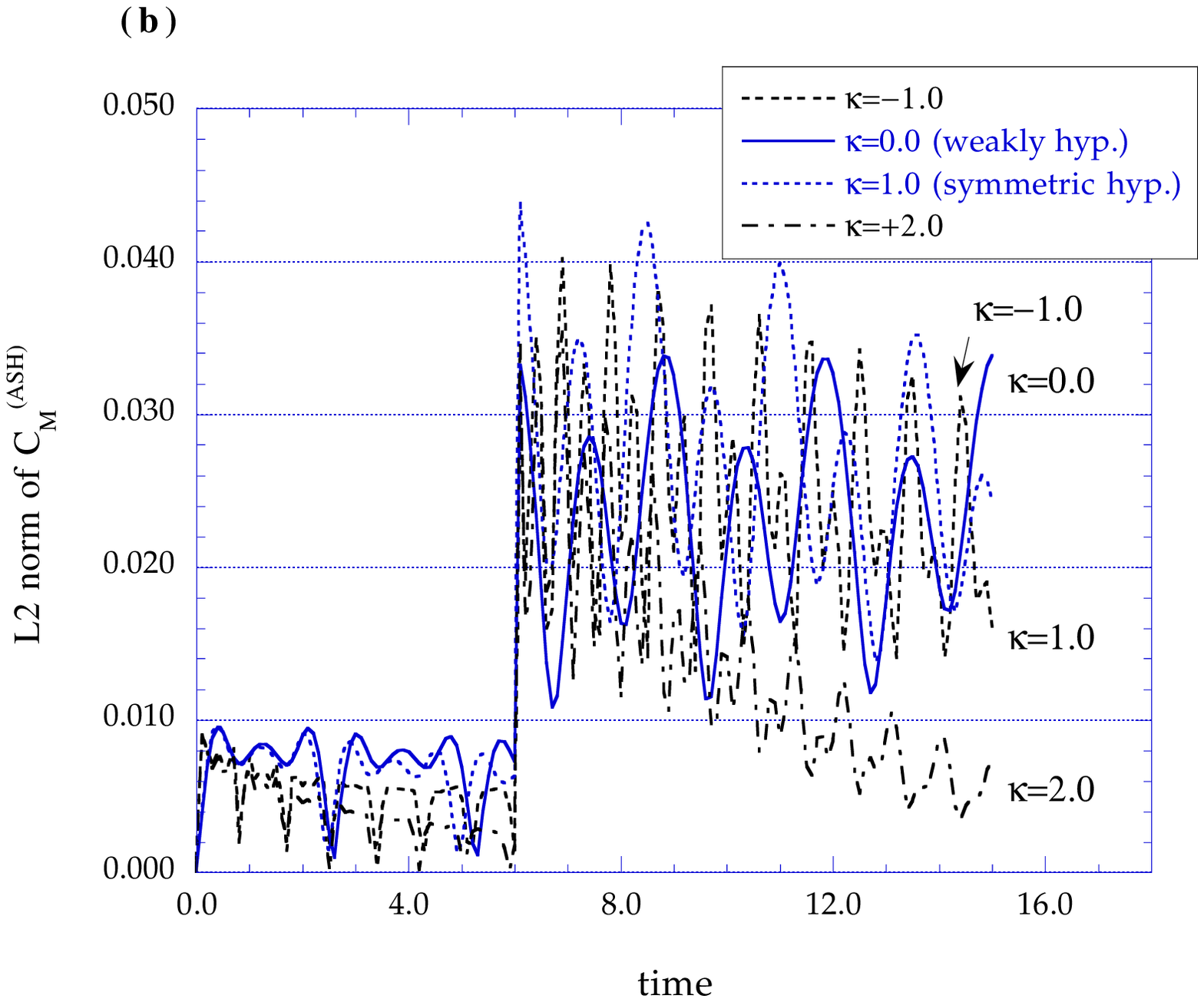} }
\end{picture}

\caption[fig-acc]{\small 
Demonstration of the adjusted system in the Ashtekar equation.
An artificial error term was added
at $t=6$, in the form of ${\cal A}^2_y \rightarrow 
{\cal A}^2_y (1+ \mbox{ error})$, where
error is +20\% as before.
Fig. (a) and (b) are L2 norm of the Hamiltonian
constraint equation, ${\cal C}_H$, and momentum constraint equation,
${\cal C}_{Mx}$, respectively.
The solid line is the case of $\kappa=0$, that is the case of ``no adjusted"
original Ashtekar equation (weakly hyperbolic system).
The dotted line is for $\kappa=1$, equivalent to the symmetric hyperbolic
system.  We see other line  ($\kappa= 2.0$) shows better performance
than the symmetric hyperbolic case.
}
\label{errwave2}
\end{figure}

We here 
examine how the adjusted multipliers contribute to the system's stability
\cite{Paper2}.
In Fig.\ref{errwave2}, we show the results of this experiment.
We plot the violation of the constraint equations both ${\cal C}_H$
and ${\cal C}_{Mx}$.
An artificial error term was added in the same way as above.
The solid line is the case of $\kappa=0$, that is the case of ``no adjusted"
original Ashtekar equation (weakly hyperbolic system).
The dotted line is for $\kappa=1$, equivalent to the symmetric hyperbolic
system.  We see other line ($\kappa=2.0$) shows better performance
than the symmetric hyperbolic case.

\section{Why adjusted equations have better performance?} \label{appC}
The remaining question is: 
why we can get the better performance by adding
constraint terms in the dynamical equations?
The added terms are basically {\it error} terms during the evolution
for its original dynamical equations.
Nevertheless, these terms improve the accuracy of the evolution.
This question applies to both higher level of hyperbolicities and
also for our proposing ``adjusted systems".
Here we introduce our plausible explanation schematically. 
The detail explanations and numerical experiments are
in \cite{Paper2}.

Suppose we have constraint equations, ${\cal C}_1 \approx 0, \,
{\cal C}_2 \approx 0, \cdots$,  in a system.
We, normally,  monitor the error of the evolution by evaluating these
constraint equations on the each constant-time hypersurface.
Such monitoring, on the other hand, can be performed also by
checking the evolution equations of the constraint, which we denote
constraint propagation equations (cf. \cite{Fri-con}).
If this set of constraint propagation equations could be written in
a first order form, then we may predict the evolution behavior by
its characteristic matrix, $M$, which is expressed by
\begin{equation}
\partial_t \left( \matrix{{\cal C}_1 \cr {\cal C}_2 \cr \vdots }\right)
\simeq M \partial_i
\left( \matrix{{\cal C}_1 \cr {\cal C}_2 \cr \vdots}\right),
\label{c1}
\end{equation}
where $\simeq$ means we are extracting only the characteristic part.

The idea here is to estimate the eigenvalues of the
characteristic matrix, $M$, after we took the Fourier transformation
on the both sides of (\ref{c1}). 
\been
\item 
Clearly, if all the real part of the eigenvalues are negative,
then all constraints decays to zero along to the system's evolution.
\item Alternatively, if all the eigenvalues are pure-imaginary,
then all constraints evolve similar to wave equations, and 
diffusion due to differenciation may preserve from its unstable evolution. 
\enen
These guidelines are supported by von Neumann's stability analysis. 
For the adjusted Ashtekar equations, the eigenvalues of the
amplicication matrix in von Neumann's method for the set of 
$({\cal C}_H, {\cal C}_{Mi}, {\cal C}_{Ga})$ are 
$\{ 1, 
1-\nu^2 \kappa \sin\theta
\pm i\nu \kappa |\sin\theta| (\mbox{multiplicity~2~each}),
1-\nu^2(1-2\kappa)^2 \sin^2\theta
\pm i \nu |1-2\kappa| |\sin\theta|
\}$, where $\nu=\Delta t/ \Delta x$, $\theta$ is the parameter for 
the frequency in grid spacing domain, and $\kappa$ is the adjusted parameter
(\S 2.4).  We observe that von Neumann's necessary condition for the stability 
can be obtained in larger non-zero $\kappa$. 

In \cite{Paper2}, we show that such a case can be obtained
by adding `adjusted terms' both for Ashtekar's and Maxwell's systems.
There we also show examples of unstable evolution by choosing adjusted
terms which produce positive eigenvalues of $M$.  

In this point, we can say that adjusted terms are responsible for
obtaining the stable and/or accurate evolution system, and this is a 
way to control the stability of simulation, which effects more than the
system's hyperbolicity. 

\section{Concluding remarks}
We presented 
numerical comparisons on the accuracy/stability of the systems, 
mainly between three mathematical levels
of hyperbolicity:
weakly hyperbolic, strongly hyperbolic,
and symmetric hyperbolic systems.
We apply Ashtekar's connection formulation, because this is the only
known system in which we can compare three hyperbolic levels with the same
interface. 
We also tested two another approaches with a purpose of constructing 
an ``asymptotically constrained" system, 
``$\lambda$-system" and ``adjusted system",  
which can be robust against perturbative
errors for the free evolution of the initial data. 

We may conclude that higher levels of hyperbolic formulations
help the numerics more, though its differences are small \cite{hern}.
Two other approaches work as desired.  These show quite good 
performance than those of the 
original and its symmetric hyperbolic form. 
This indicates, in turn, that the symmetric hyperbolic system is not always
the best for controlling accuracy or stability.

The reason why higher hyperbolicity shows better stability may be 
explained by the eigenvalue analysis of the 
propagation equation of the constraints. 
Our two guidelines, negative-real eigenvalues or pure-imaginary eigenvalues, 
give us clear indications whether
the constraints decay (i.e. stable system) or not for 
perturbative errors, at least to all our numerical experiments. 

We think these results open a new direction to numerical relativists for future
treatment of the Einstein equations.  

To conclude, we are glad to announce that Ashtekar's connection variables
have finally been applied in numerical simulations.  
This new approach, we hope, will contribute to understanding further 
of gravitational physics, and will open a new window for peeling off
interesting non-linear natures together with a step to numerical treatment
of quantum gravity.

\section*{Acknowledgements}
HS appreciates helpful comments by
  Abhay Ashtekar, Jorge Pullin, Douglas Arnold and L. Samuel Finn,
and the hospitality of the CGPG group. 
Numerical computations were performed using machines at CGPG.
This work was supported in part by the NSF grants PHYS98-00973, 
and the Everly research funds of Penn State. 
HS was supported by the Japan Society for the Promotion of Science
as a research fellow abroad.  HS appreciates the travel support 
from the LOC of the JGRG workshop.

\end{document}